\documentclass[
superscriptaddress,
showkeys,
preprint,			
bibnotes,
amsmath,amssymb,
aps,
prb, 
]
{revtex4-2}						

\usepackage{color}
\usepackage[normalem]{ulem}
\usepackage{graphicx}
\usepackage{bm}
\usepackage{xspace}
\usepackage{setspace}
\usepackage{float}

\newcommand{\nips}{$\mathrm{NiPS}_3$\xspace}

\newcommand{\mipx}{$\mathrm{MPX}_3$\xspace}
\newcommand{\znps}{$\mathrm{Ni_{1-x}Zn_{x}PS}_3$\xspace}
\newcommand{\cdps}{$\mathrm{Ni_{1-x}Cd_{x}PS}_3$\xspace}
\newcommand{\nipse}{$\mathrm{{NiPS}_{3-x}Se_{x}}$\xspace}

\newcommand{\nicl}{$\mathrm{{NiCl}_{2}}$\xspace}

\newcommand{\tn}{$\mathrm{T_N}$\xspace}
\newcommand{\sing}{$\mathrm{^1}E$\xspace}
\newcommand{\trip}{$\mathrm{^3}A_2$\xspace}
\newcommand{\asym}{$\mathrm{^1}A_1$\xspace}
\newcommand{\tsym}{$\mathrm{^3}T_1$\xspace}
\newcommand{\ttsym}{$\mathrm{^3}T_2$\xspace}

\begin{document}   

\title{Crystal field tuned spin-flip luminescence in $\mathrm{NiPS}_3$}

\author{L\'eonard Schue}
\affiliation{%
Laboratoire de Physique de la Matière Condensée, CNRS, Ecole Polytechnique, Institut  Polytechnique de Paris, 91120 Palaiseau, France 
}%
\author{Nashra Pistawala}
\affiliation{%
Department of Physics, Indian Institute of Science Education and Research, Pune, Maharashtra-411008, India
}%
\author{Hebatalla Elnaggar}
\affiliation{%
Sorbonne Université, Muséum National d'Histoire Naturelle, UMR CNRS 7590, IRD, Institut de Minéralogie, de Physique des Matériaux et de Cosmochimie, IMPMC, 75005 Paris, France
}%
\author{Yannick Klein}
\affiliation{%
Sorbonne Université, Muséum National d'Histoire Naturelle, UMR CNRS 7590, IRD, Institut de Minéralogie, de Physique des Matériaux et de Cosmochimie, IMPMC, 75005 Paris, France
}%
\author{Christophe Bellin}
\affiliation{%
Sorbonne Université, Muséum National d'Histoire Naturelle, UMR CNRS 7590, IRD, Institut de Minéralogie, de Physique des Matériaux et de Cosmochimie, IMPMC, 75005 Paris, France
}%
\author{Johan Biscaras}
\affiliation{%
Sorbonne Université, Muséum National d'Histoire Naturelle, UMR CNRS 7590, IRD, Institut de Minéralogie, de Physique des Matériaux et de Cosmochimie, IMPMC, 75005 Paris, France
}%
\author{Fausto Sirotti}
\affiliation{%
Laboratoire de Physique de la Matière Condensée, CNRS, Ecole Polytechnique, Institut  Polytechnique de Paris, 91120 Palaiseau, France
}%
\author{Yves Lassailly}
\affiliation{%
Laboratoire de Physique de la Matière Condensée, CNRS, Ecole Polytechnique, Institut  Polytechnique de Paris, 91120 Palaiseau, France
}%
\author{Fabian Cadiz}
\affiliation{%
Laboratoire de Physique de la Matière Condensée, CNRS, Ecole Polytechnique, Institut  Polytechnique de Paris, 91120 Palaiseau, France
}%
\author{Luminita Harnagea}
\affiliation{%
Department of Physics, Indian Institute of Science Education and Research, Pune, Maharashtra-411008, India
}%
\affiliation{%
I-HUB Quantum Technology Foundation, Indian Institute of Science Education and Research, Pune 411008, India
}
\author{Abhay Shukla}
\email{abhay.shukla@sorbonne-universite.fr}
\affiliation{%
Sorbonne Université, Muséum National d'Histoire Naturelle, UMR CNRS 7590, IRD, Institut de Minéralogie, de Physique des Matériaux et de Cosmochimie, IMPMC, 75005 Paris, France
}%

\begin{abstract}
Layered magnetic materials potentially hold the key to future applications based on optical control and manipulation of magnetism. \nips, a prototype member of this family, is antiferromagnetic below 155 K and exhibits sharp photoluminescence associated to a transition between a triplet ground state and a singlet excited state. The nature of the luminescent transition is a matter of current debate and so is an eventual fundamental link of this excitation to magnetism. Here we provide answers through experiments and calculations. We fabricate samples with metal and ligand substitutions which alter the Néel transition temperature and measure the effects of these changes on the temperature dependent photoluminescence. We perform crystal field and charge transfer multiplet calculations to explain the origin of the excitation and identify the effects of the magnetic ground state on its properties. These measurements and calculations provide a comprehensive explanation for the observed properties and a template for finding similar materials exhibiting spin-flip luminescence.

\begin{description}
\item[DOI] 
\hspace{8cm} 
\textbf{PACS numbers:}
\end{description}
\end{abstract}

\maketitle  

\section{Introduction}
Magnetic van der Waals materials \cite{Wang2018,Basnet2024} are ideal for investigating magnetism and magnetic transitions in two dimensions (2D). They provide a platform to study the coupling of magnetic degrees of freedom with optical and electronic properties for future applications in optoelectronic, spintronics and quantum information. Because of the layered nature of these materials devices of hetero-structures can be envisaged. Metal phosphorous trichalcogenides (\mipx) with (M = Mn, Fe, Co, Ni and X = S, Se,Te) form one such family of magnetic layered materials and \nips has been the focus of much recent work fueled by the observation by Kang \emph{et al}.~\cite{Kang2020} of a sharp photoluminescent (PL) excitation at 1.475 eV with several intriguing characteristics. This PL peak is visible  below roughly 100 K. Since the antiferromagnetic transition temperature (\tn) is about 150 K the exciton was assumed to be of magnetic origin. At low temperature its width was found to be surprisingly sharp ($<$ 1 meV) which prompted claims of a coherent bosonic excitation. A smaller twin of the 1.475~eV luminescence about 2 meV higher in energy was observed and absorption measurements also showed a broader peak at an energy 23 meV higher which Kang \emph{et al.} identified as a double magnon sideband. They also made resonant inelastic X-ray scattering (RIXS) measurements which are less constrained by selection rules. They concluded that the 1.475 eV emission, also visible by RIXS along with several other excitations, is a transition from a singlet state to a triplet state. They inferred these to be hybridized Zhang-Rice states of Ni and S orbitals with efficient charge transfer between the two. This interpretation of a coherent, delocalized excitation was also put forward by other works~\cite{Hwangbo2021} which underlined other properties such as a high degree of linear polarization of the emitted light~\cite{Wang2021}, its link to the antiferromagnetic transition as deduced through the substitution of non-magnetic Cd for Ni~\cite{Kim2023} or through the splitting with a high in-plane magnetic field~\cite{Wang2024}. The picture that emerged was of a robust collective excitation with a possible unusual coherent ground state and a direct link to magnetism allowing for coupling between local spin structure and light emission. This interpretation was subsequently questioned by several works which instead proposed a d-d excitation localized on the Ni ion~\cite{Ergecen2022,Dirnberger2022,Jana2023,He2024}. The role of magnetic long-range order was also discussed and a fundamental link between long-range magnetism and the 1.475 eV emission was put into doubt by some authors who invoked the symmetry breaking aspect~\cite{Ergecen2022,Jana2023} of the antiferromagnetic ground state as a factor in making the emission bright.
These intriguing aspects are still debated, awaiting a conclusive explanation. We answer these questions with experiments on metal and ligand substituted samples and by making crystal field and charge transfer multiplet calculations. We find that the singlet-triplet luminescence originates in a localized excitation of Ni 3d crystal field levels known as spin-flip luminescence in photochemistry~\cite{Kitzmann2022}. The brightness of this excitation and its manifestation as a luminescence signal is sensitive to the value of the crystal field and its screening and although its properties are  influenced by the magnetic ground state, it is not as such, a magnetic excitation. However such materials, where magnetism and optical excitations are coupled, are of great interest for a variety of future applications such as opto-spintronics~\cite{Nmec2018}, optical switching of magnetization~\cite{Zhang2022} and optical manipulation of spin~\cite{Bayliss2020}.

\section{Results}
\begin{figure}[H]
\centering
\includegraphics[width=0.8\textwidth]{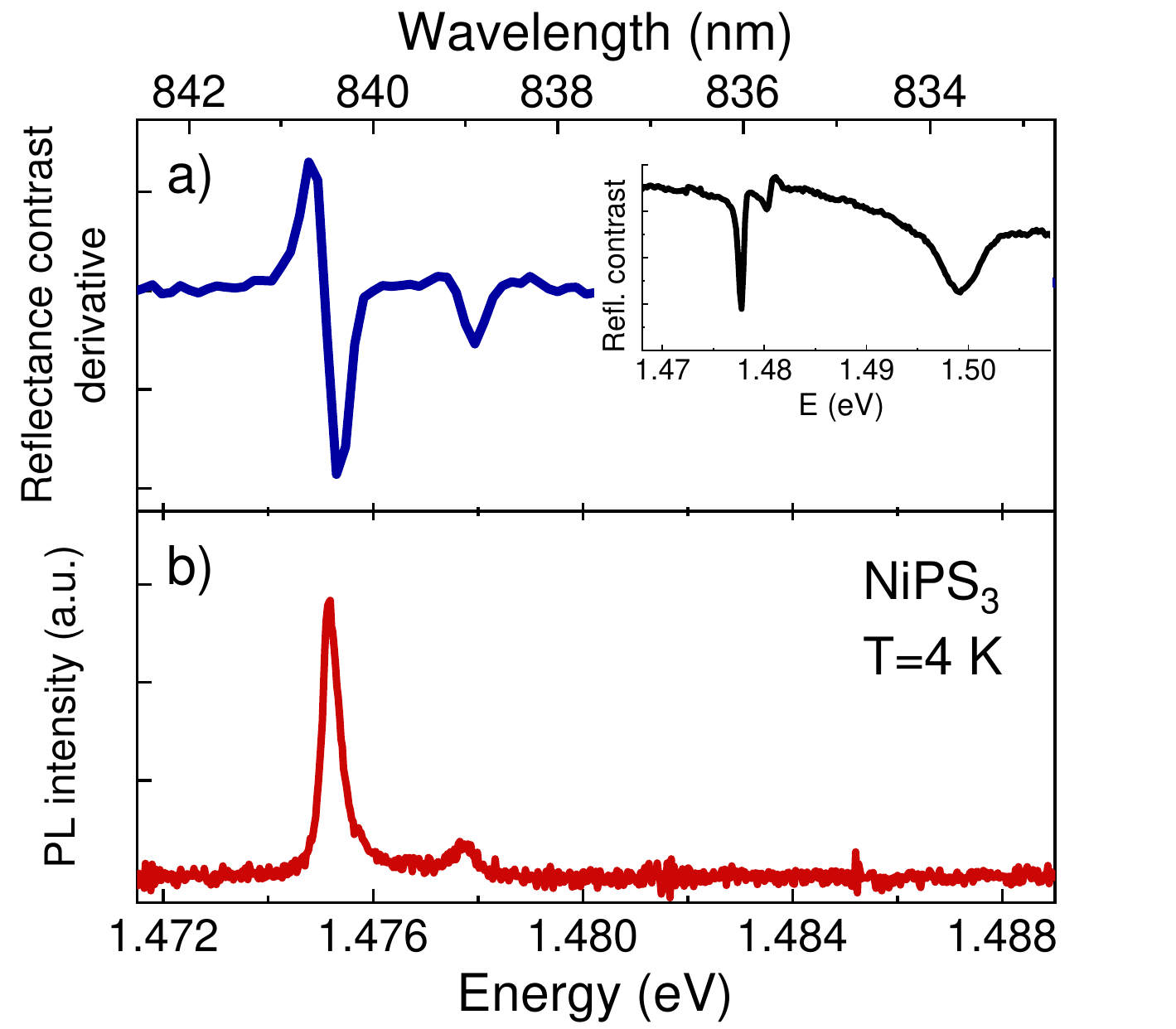}					 
\caption{\nips at 4 K: a) Reflectance contrast (inset) and its derivative as a function of incident energy.  b) Photoluminescence over the same energy range.
}
\label{fig1}
\end{figure}

We measured magnetization and PL as a function of temperature in pure \nips and in the substituted compounds \znps and \nipse. Technical details can be found in the Experimental Section.  These measurements allow us  to conclusively establish the physical origin of the luminescence, probe the link with magnetism and interpret, and explain existing results in literature. In Fig. \ref{fig1}a we show the change in reflectance as a function of photon incident energy at 4 K for a \nips sample while the PL is shown in Fig.  \ref{fig1}b. While the excitations at 1.475 and 1.477 eV are visible both in absorption (dips in reflectance in the inset of Fig. \ref{fig1}a) and in PL measured with a photon energy of 1.746 eV, the 1.498 eV excitation is not radiative.

One of the questions concerning the origin of these excitations is their link to magnetism. Earlier work~\cite{Kang2020,Hwangbo2021,Wang2021} highlighted the appearance of the 1.475 eV excitations below the \tn of \nips as proof of a fundamental link. Indeed, Kim \emph{et al}.~\cite{Kim2023} substituted non-magnetic Cd for magnetic Ni and found that  10\% Cd substitution depressed \tn below 115K and simultaneously completely suppressed PL, confirming this conclusion. Elemental substitution and its effect on \tn in \mipx has been studied in detail \cite{Basnet2024}. While substitution of a non-magnetic metal for Ni effectively suppresses the \tn in \mipx compounds, ligand substitution has varying effects depending on the compound. In \nips, substituting S by Se results in an enhancement of \tn. This is ascribed to the greater orbital overlap between metal and ligand through more extended Se 4p orbitals, enhancing the superexchange mediated interaction between Ni ions.

\begin{figure}[H]
\centering
\includegraphics[width=0.8\textwidth]{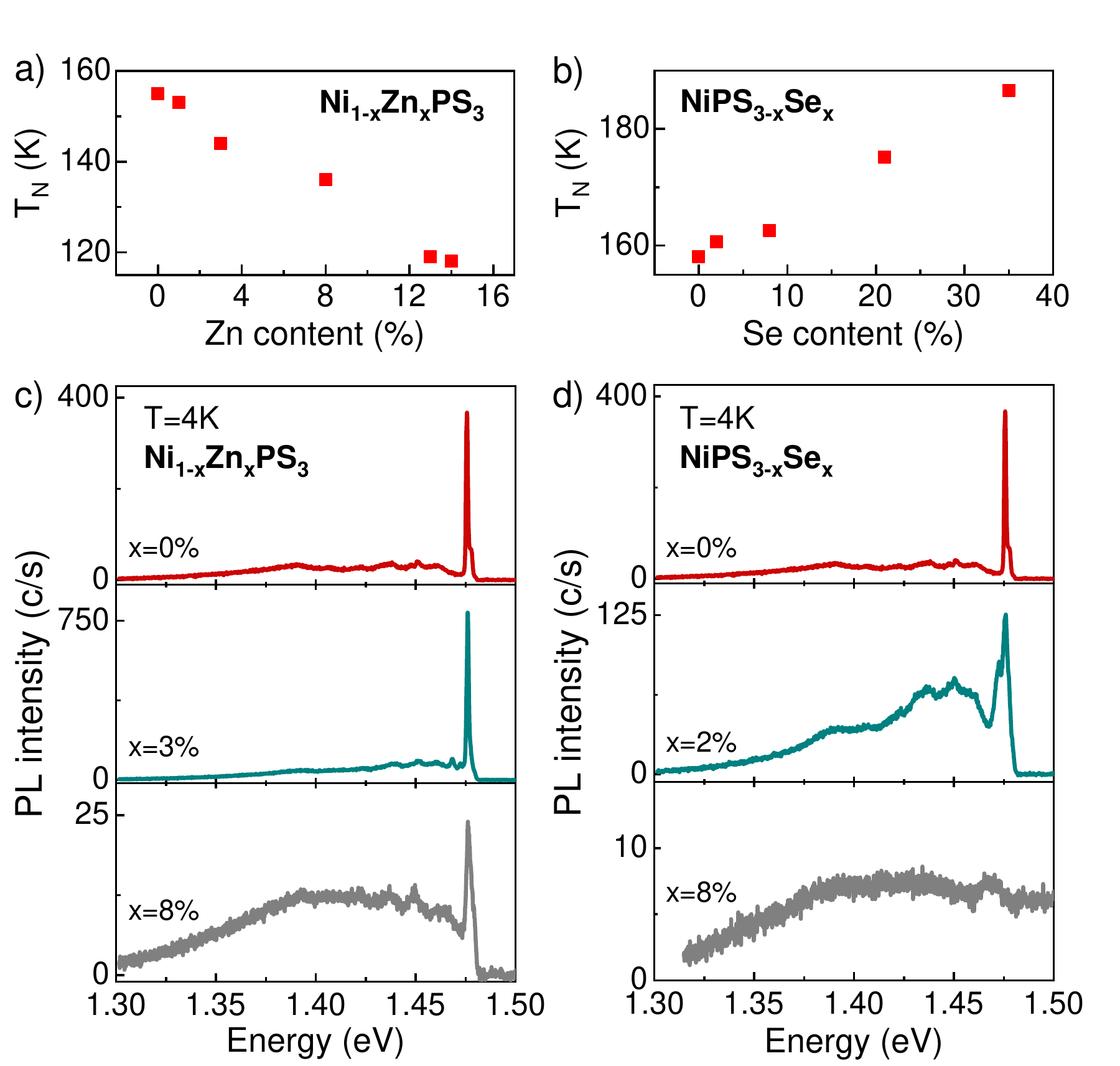}						
\caption{The effect of metal and ligand substitution. \tn as a function of a) Zn substitution for Ni and b) Se substitution for S. PL as a function of c) Zn substitution for Ni and d) Se substitution for S.
}
\label{fig2}
\end{figure}

\begin{figure}[h!]
\centering
\includegraphics[width=0.8\textwidth]{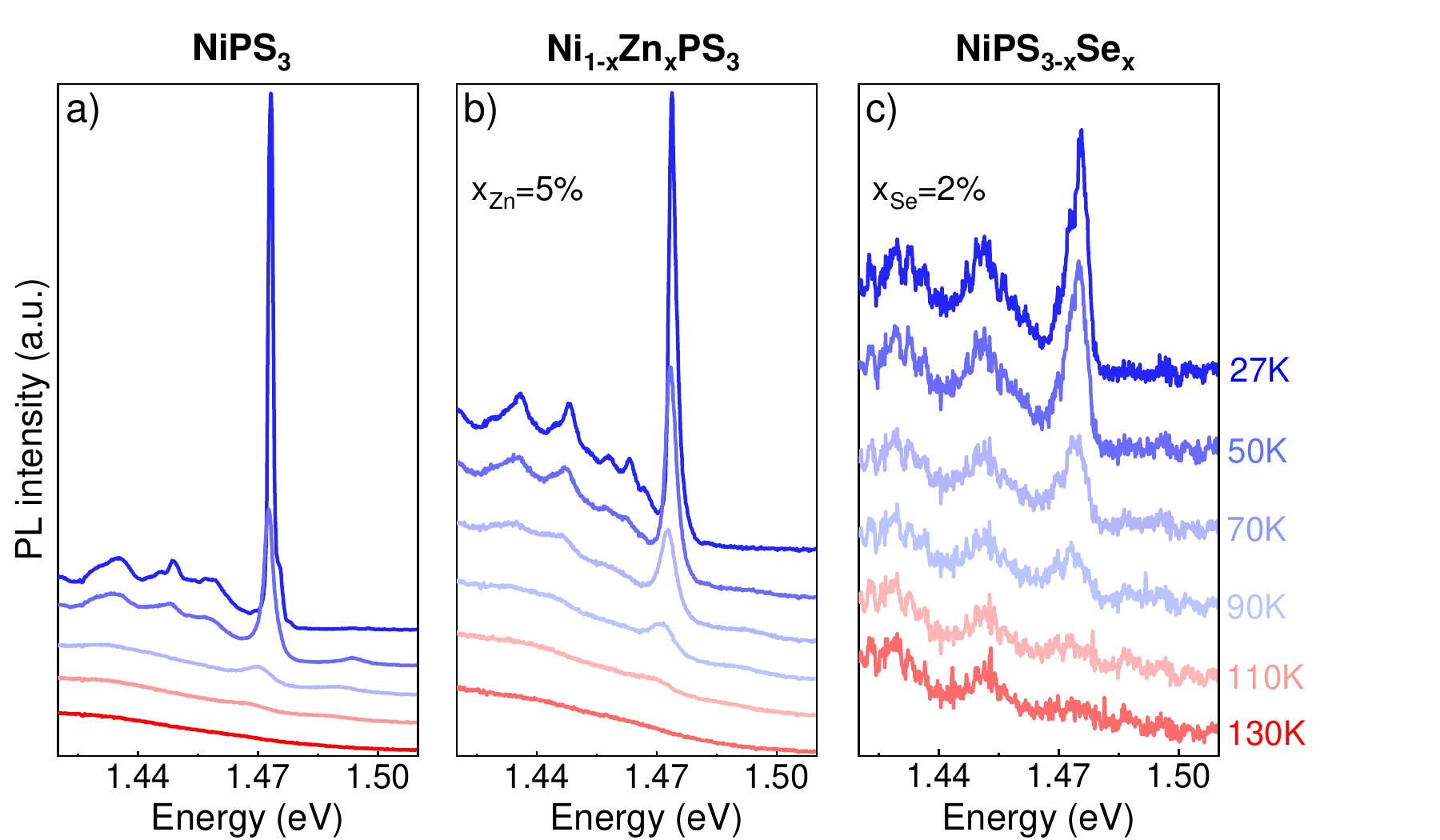}						
\caption{Temperature dependence of PL with elemental substitution. a) \nips (the 70 K spectrum is absent) b) 5\% Zn substitution for Ni c) 2\% Se substitution for S
 }
\label{fig3}
\end{figure}

Our Zn~\cite{Pistawala2024} and Se substituted samples confirm to this \tn dependence on substitution as shown in Fig. \ref{fig2}a) and b). However the 4 K PL measurements  give surprising results. Although the intensity of the the 1.475 eV feature decreases and the peak broadens with Zn substitution in Fig. \ref{fig2}c), it is still clearly visible (here shown for 8\% Zn substitution, however we have measured a clear peak upto the maximum substitution reached of 14\% as shown in Fig. S5). This is in contrast to the complete suppression of PL at 10\% Cd substitution reported in \cite{Kim2023}. While a decrease in PL intensity and broadening could be expected because of the disorder introduced by the substitution of a non-magnetic ion, the different trends for Zn and Cd substitution are intriguing. Even more striking are the effects of Se substitution shown in Fig. \ref{fig2}d). For 8\% Se substitution the 1.475 eV feature is completely suppressed while \tn is actually 10 K higher than in \nips.
 In Fig. \ref{fig3} we show the temperature dependent PL for the pure as well as  Zn (5\%) and Se (2\%) substituted samples. Despite a difference in \tn of more than 25K between these samples there is no apparent correlation between \tn and the temperature ($\approx$ 100 K) below which the 1.475 eV feature appears. These results unequivocally indicate that the exciton does not have a magnetic origin although its properties could well be influenced by the magnetic ground state.

\section{Discussion}
\begin{figure}[h!]
\centering
\includegraphics[width=0.8\textwidth]{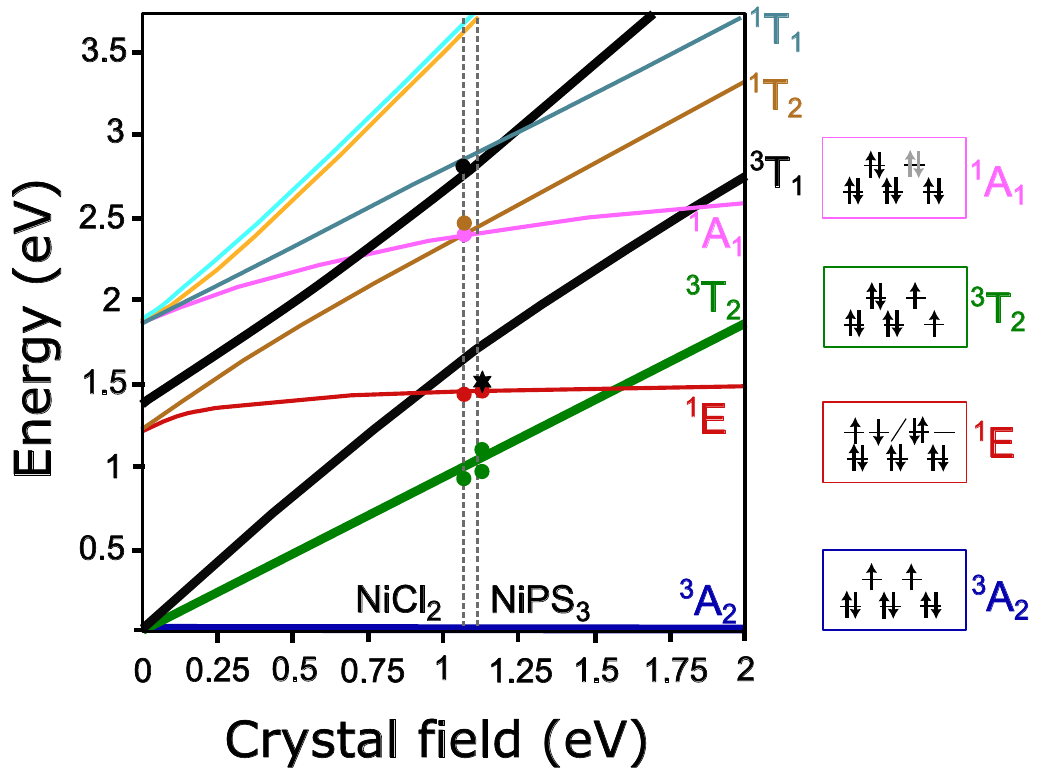}						
\caption{Tanabe Sugano diagram for the $\mathrm{d^8}$\xspace $\mathrm{Ni^{2+}}$\xspace ion in an octahedral crystal field. The dashed vertical lines represent a certain crystal field value which characterizes either \nips or \nicl. The intersection with the crystal field energy levels indicated by their symmetry establishes the excited states of the ion for the material. On the right the spin configurations and term symbols of some levels are specified.
}
\label{fig4}
\end{figure}

Light emission and photoactivity are essential elements of devices and applications. The photoactivity of compounds of transition metals is widely studied for applications as diverse as photocatalysis~\cite{Twilton2017} or modern light emitting devices~\cite{Bizzarri2018}. In  molecular complexes of transition metals most emissions are due to transitions between metal-ligand or ligand-ligand charge transfer states. A smaller subset is due to spin-flip luminescence centered on the metal ion and arising from singlet to triplet transitions which are, in general, both spin and symmetry forbidden but still observable under conditions which can be quantified~\cite{Kitzmann2022}. These emissions from metal centered transitions can be studied using a well-known analysis tool known as  the Tanabe Sugano (TS) diagram~\cite{Tanabe1954,Tanabe1954b} which provides the splitting of the transition metal ion energy levels as a function of the strength of the crystal field. The TS diagram is of generic nature for a given d-shell filling ($\mathrm{d^8}$\xspace for the $\mathrm{Ni^{2+}}$\xspace ion). It does not account for distortions from ideal symmetry or for charge transfer states, both of which need more detailed calculations for proper characterization. However as we shall see in Fig. \ref{fig4}, it allows a useful estimation of energy levels and transitions and can account summarily for changes introduced by chemical bonding through Racah parameters which characterize electronic repulsion and intervene in the determination of the energy levels and their dependence on the crystal field. In general the degree of covalency cannot be directly mapped onto the crystal field strength. However it provides a useful indication, in particular in substituted \nips where increase of covalency is accompanied by an increase in the bond length (Ni-S bond length = 2.44 \AA \xspace while Ni-Se bond length = 2.54 \AA ~\cite{Gu2019}).

In Fig. \ref{fig4} we show the $\mathrm{d^8}$\xspace TS diagram in an octahedral crystal field~\cite{Haverkort2012}. 
The ground state (\trip symmetry) is a spin 1 state. The thin lines correspond to levels with a different spin from the ground state (here spin 0 singlets). Some levels show little variation (\sing and \asym) with the crystal field because the corresponding electronic distribution is similar to the ground state and transitions to these states, `nested' with the ground state in configuration space, are symmetry forbidden. In particular the transition to \sing is both spin and symmetry forbidden. At high crystal field strength these are the lowest levels. Others, shown with thick lines (\tsym and \ttsym for example), are spin allowed and may be translated with respect to the ground state in configurational space if their electronic distribution is different. Spin-flip transitions (\trip to \sing or \asym)  are optically sharp while transitions to spin and symmetry allowed multiplets are generally broad~\cite{Kitzmann2022}. In Fig. \ref{fig4} we correlate peaks (dots) from published RIXS measurements to find the crystal field for \nips ~\cite{Kang2020,He2024} and \nicl ~\cite{Occhialini2024} and also plot the energy of the 1.475 eV exciton (star). With this assignment the exciton is identified as the spin-flip luminescence after photoexcitation from the \trip ground state to the \sing state. As mentioned earlier, several works~\cite{Kang2020,Hwangbo2021,Wang2021,Kim2023,Wang2024} have instead classified this as a triplet-singlet Zhang-Rice charge transfer orbital transition. Others opt for an unspecified d-d transition~\cite{Ergecen2022,Dirnberger2022,Jana2023} or one from the \asym state to the \trip state~\cite{He2024}. 

\begin{figure}[h!]
\centering
\includegraphics[width=0.8\textwidth]{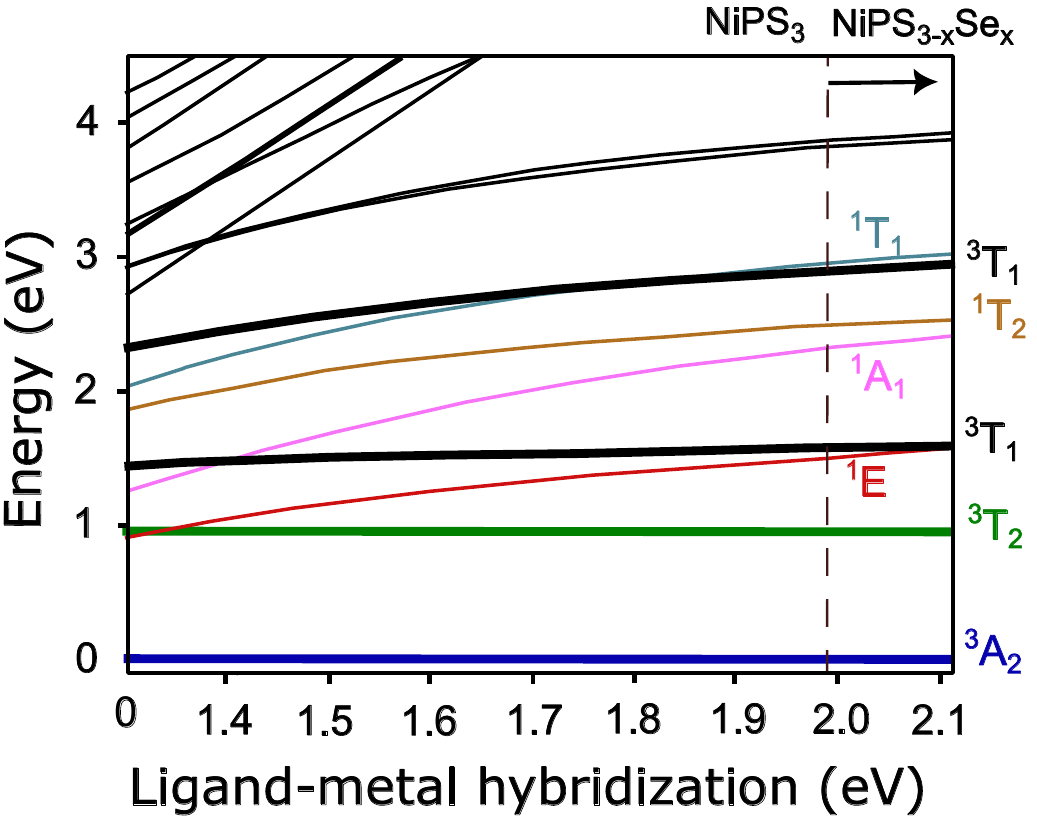}						
\caption{Energy-level multiplet diagram in the charge-transfer model as a function of the ligand-metal hybridization which includes  V$_{eg}$, V$_{t2g}$ and $\Delta$  to capture the covalency as described relation in the Experimental Section. On the right, the term symbols of some levels are specified.
}
\label{fig5}
\end{figure}

For a more rigorous approach we perform charge transfer multiplet (CTM) energy-level calculations which include the charge transfer from the ligand orbitals to the Ni ion. In other words, the electronic states of Ni can  be described by the configuration interaction between three configurations which are  d$^8$ + d$^9$\underline{L} + d$^{10}$\underline{L}$^2$. In this model, three extra parameters come in  to play which are the hybridization strength between the ligand and metal orbitals (V$_{eg}$ and V$_{t2g}$) and the charge transfer gap between the configurations ($\Delta$). The details about the model parametrization are given in the Experimental Section. Fig. \ref{fig5} shows the energy-level diagram as a function of the ligand-metal hybridization (V$_{eg}$ and V$_{t2g}$) where we find that the electronic structure of NiPS$_3$ is well captured with V$_{eg} \approxeq$ 2 eV (vertical dashed line). This agrees well with the energy TS diagram where the ground state is \trip as expected. We confirm that our assignment of the 1.475 eV exciton to \trip ~-\sing spin-flip luminescence remains valid. With the TS diagram we estimated similar crystal fields in \nips and \nicl, an estimation which is borne out by our CTM calculations and those of Occhialini \emph{et al}. \cite{Occhialini2024} as well as by the similar Ni-S (2.44 \AA) and Ni-Cl (2.43 \AA) bond lengths~\cite{Gu2019,Ouvrard1985,Brik2015}.

We now account for the various characteristics of this luminescence. Spin-flip luminescence, while both spin and symmetry forbidden, may be observed through a symmetry breaking mechanism often identified in the case of \nips as the antiferromagnetic ordering~\cite{Ergecen2022,Jana2023}. Other possible mechanisms are coupling to symmetry breaking thermal vibrations or spin-orbit coupling of the \sing excited state to the nearby spin-allowed \tsym state (see Fig. \ref{fig4}) as proposed by Pisarev~\cite{Pisarev1966}. Coupling to phonons would however imply the opposite temperature dependence to the one observed. The proximity to the \tsym state can also explain an important observation which is the intriguing suppression of the luminescence with Se substitution although \tn is enhanced. The increase in \tn is due to increased superexchange and orbital overlap through the bigger size of the Se atom (Ni-Se bond length = 2.54 \AA ~\cite{Gu2019}) and the stronger covalent character of the bond. In Fig. \ref{fig5}, where energy levels are plotted as a function of ligand-metal hybridization, this implies that the vertical \nipse line moves to the right with respect to the \nips position (increasing covalent character). In Fig. \ref{fig4} for the isolated Ni ion this corresponds to a translation to the left towards a lower crystal field strength. In both cases the \sing ~- \tsym crossing is approached and spectral weight is transferred to the \trip ~- \tsym transition which may even become lower in energy. This transition is generally characterized by a broad lineshape and enhanced non-radiative processes, suppressing luminescence. 
\\
The requirement for the spin-forbidden level to be reasonably lower in energy than the spin-allowed level to observe spin-flip luminescence is in fact an important design criterion for tuning crystal field strength in luminescent complexes \cite{Kitzmann2022} and importantly, remains valid for designing layered magnetic materials. It is also the probable reason for  luminescence suppression in \cdps with respect to \znps because of the contrasting effect of these two substitutions on covalency, screening and eventually on the effective crystal field which tunes the relative positions of these energy levels.
The narrow luminescence line (FWHM of 475 $\mathrm{{\mu}eV} $) of \nips at 4K (Fig. \ref{fig1}b) broadens as the temperature increases and vanishes close to and above \tn. These observations have prompted suggestions of an essential link to magnetism. Similar d-d transitions in crystals of transition metals have often been measured with optical absorption in the past (including in \nips ~\cite{Banda1986}). $\mathrm{{NaNiF}_{3}}$\xspace \cite{Pisarev1966} and $\mathrm{{KNiF}_{3}}$\xspace \cite{Pisarev1968} show  sharp absorption lines from \trip to \sing transitions which markedly decreases in intensity for the latter across \tn. Similar observations have been made in $\mathrm{{KMnF}_{3}}$\xspace \cite{Ferguson1966} and $\mathrm{{NaCrO}_{2}}$\xspace \cite{Elliston1974}. In several Cr compounds Schmidt \emph{et al}.~ \cite{Schmidt2013} measure d-d transitions (linewidths well below 1 meV at low temperature) which broaden and disappear at or above \tn while Multian \emph{et al}.~\cite{Multian2025} in recent measurements in $\mathrm{{CrPS}_{4}}$\xspace also observe a similar temperature dependence for luminescent d-d excitations. 
\\
Tanabe \emph{et al}.~\cite{Tanabe1965} have argued for a dipole-allowed character relaxing the symmetry constraint for the transition and which could appear through interaction with magnons  in the antiferromagnetic phase. Magnetic fluctuations at higher temperature and certainly above \tn would contribute to a damping mechanism causing intensity suppression. Inversely, the strikingly narrow low temperature luminescence in \nips could originate in the decay to the spin ordered low temperature ground state, similar to the line narrowing of excitons induced by high magnetic fields which align spins in diluted magnetic systems~\cite{Bacher2001}. Instead of a strict correlation with \tn the picture that emerges is that of sharp transitions (luminescent or not) at low temperature and intensity suppression and broadening at high temperature.
\\
The \trip ~- \sing spin-flip luminescence corresponds to a spin triplet ground state and an orbital doublet excited state (see Fig. \ref{fig4}) of which both \sing microstates are populated and thus prone to Jahn-Teller splitting. We propose that  the observed doublet at 1.475 and 1.477 eV (Fig. \ref{fig1}) corresponds to this splitting in the absence of a magnetic field. The splitting of the luminescence observed with high magnetic field \cite{Wang2021,Jana2023} however, is clearly Zeeman splitting of the \trip ground state~\cite{Bayliss2020} corresponding to the spin orientations parallel and anti-parallel to the applied field. The absorption (non-luminescent) peak at 1.498 eV has been attributed  to a double magnon sideband \cite{Kang2020} of the luminescence but the energy difference does not correspond to the measured (roughly 80 meV) double magnon energy~\cite{Kang2020,He2024}. On the other hand \nips has a strong IR-active phonon with an energy of 23 meV \cite{Sourisseau1983}, corresponding to the exact energy difference between the two peaks. The 1.498 eV peak is a phonon sideband of the luminescence, accounting for the high intensity of this peak as well as its non-radiative nature. This also contradicts the claim~\cite{Hamad2024} which identifies both the 1.475 eV and the 1.498 eV excitations as singlet polarons and notably predicts radiative decay for the 1.498 eV peak.

\section{Conclusion}
In this work we answered open questions concerning the origin and properties of sharp low temperature photoluminescence in the prototype antiferromagnetic layered 2D material \nips which has been the focus of many recent studies. We investigated a series of compounds with partial substitution of the Ni or the S atom by Zn and Se respectively using photoluminescence and magnetization measurements as a function of temperature. We calculated the 3d energy levels of the Ni ion in an octahedral ligand field for the bare ion and the Ni-S octahedron taking into account charge transfer, screening and the covalency which ensues. We found that this luminescence which depends critically on the ligand field and on Zn and Se substitution, corresponds to a singlet (non-magnetic \sing excited state) to triplet (magnetic \trip ground state) crystal field transition as opposed to a transition between charge transfer states.  We found that Zn substitution for Ni degrades  \tn but does not affect PL in a critical way while Se substitution increases \tn but intriguingly destroys PL, contradicting a direct link with magnetism. Spin-flip luminescence is well-known in coordination chemistry and photochemistry as light emission from complexes of metal ions in various ligand field environments~\cite{Kitzmann2022} and under certain conditions. Several controversial or unexplained characteristics of the observed PL are explained by this interpretation. Favorable conditions exist in \nips which is a rare example of spin-flip luminescence in a crystalline non-molecular solid exhibiting a magnetic transition with long-range order.  These conditions will have to be kept in mind for designing new layered magnetic materials with similar characteristics. Interestingly, sharp polarized luminescence combined with a magnetic ground state could be a key ingredient for optical addressing of solid-state spins for future quantum information systems~\cite{Bayliss2020}.

\section{Experimental Section}
\subsection{Crystal synthesis}

Single crystals of \znps were grown through physical vapor transport, using high-purity elements enclosed in a quartz ampoule under vacuum, as precursors \cite{Pistawala2024}. The processing preliminary to crystal growth and the storage of samples were done in a glove box under an inert atmosphere. For the crystal growth of the parent compound, the source temperature was set to 750 $^{\circ}$C, while the sink temperature was set to 730 $^{\circ}$C. The growth conditions for the Zn-doped samples have been accordingly optimized to obtain a homogeneous Zn distribution. The \nipse single crystals were grown using chemical vapor transport, with iodine as the transporting agent. The growth of single crystals with less than 8\% Se-doped started directly from the chemical elements. The high-purity elements in a stoichiometric ratio and a small amount of iodine (approximately 5 $\mathrm{mg/{cm}^3}$) are loaded in a quartz ampoule, which was subsequently evacuated to  $\mathrm{{10}^{-5}}$ Torr and sealed. The ampoule was placed in a multi-zone furnace in a temperature gradient of 250 $^{\circ}$C, with the cool end set to 500 $^{\circ}$C for 10 days. For higher levels of Se-doping, initially, polycrystalline samples were prepared by solid-state reaction at 500 $^{\circ}$C, and subsequently, the homogeneous, finely powdered samples were transported as described above.
The magnetization was measured using a magnetic properties measurement system (MPMS-3) from Quantum Design. Single crystals were attached to a sample holder with GE varnish and the magnetic field applied in the ab-plan. Both zero-field cooled and field-cooled data were measured and no difference found between them. The N\'eel temperature is determined as the maximum of the first derivative of the susceptibility, d$\chi$/dT.
\subsection{Photoluminescence}
Bulk flakes of \nips, \znps and \nipse (thickness $>$ 100 nm) were mechanically exfoliated onto a $\mathrm{Si/Si{O}_2}$ substrate and cooled down to 4 K in a closed cycle cryostat. All experiments were done in back-scattering geometry using a home-built microscope equipped with a 40x objective (NA=0.6), a grating spectrometer and a CCD camera. Photoluminescence measurements were performed using a continuous Ti:Sa laser excitation with a central wavelength of 710 nm and a power ranging from 0.5 to 1 mW. A broadband radiation from a halogen lamp was used to measure the reflectance contrast $\mathrm{\Delta R = ({R}_{sample} – {R}_{substrate})/{R}_{substrate}}$. In order to highlight the weak signatures of the excitonic transitions, the derivative of the reflectance contrast $d /dE(\Delta R /R)$ in the energy range of interest is plotted in  Fig. \ref{fig1} a). PL and reflectance data shown in  Fig. \ref{fig1} were acquired using 1800 g/mm grating with a spectral resolution limit of 150 $\mu$eV while a 600 g/mm grating with a resolution limit of 600 $\mu$eV was used for the temperature- and composition-dependence measurements shown in Fig. \ref{fig2} and  Fig. \ref{fig3}.

\subsection{Tanabe Sugano and CTM calculation}
For the bare ion TS diagram we take $\mathrm{C/B=4}$\xspace where C and B are Racah integrals~\cite{Tanabe1954,Tanabe1954b} with $\mathrm{C=3000 {cm}^{-1}}$\xspace and $\mathrm{B=750 {cm}^{-1}}$\xspace \cite{Kitzmann2022,Banda1986,Nolet2006,Gonzalez2007,Brik2015}. We employ exact diagonalization calculations within charge-transfer multiplet (CTM) theory as implemented in Quanty~\cite{Haverkort2012}. The model reduces to a multi-electronic calculation of a single NiS$_6$ cluster, accounting for the Ni-3d orbitals and the corresponding symmetrized molecular orbitals from Ni 3d states with hybridization to S 3p states. The cluster has a octahedral-symmetry which is quantified through one distortion parameter (Dq). Configuration interaction calculations taking into account (i) the intra-atomic Coulomb interaction, (ii) the crystal field, (iii) charge transfer were performed. Briefly, the ligand-field interaction is given by three different terms. These are the on-site splitting on the transition-metal d-shell, the on-site splitting on the ligand symmetrized d-shell, and the hopping between the ligand d-shell and the transition-metal d-shell. The ligand states in consideration are made of linear combinations of interacting S-3p orbitals. This reduces the 36 S-3p orbitals to 10 interacting orbitals with the Ni-d orbitals. The on-site energies can be related to the charge-transfer energy ($\Delta$), and the d-d Coulomb interaction (U$_{dd}$) which includes exchange interactions following Zaanen, Sawatzky and Allen\cite{Zaanen1985}. The energy differences are referred from the center of the ligands and transition metal sites after the application of the d-electron crystal field. 
The parameters for the multiplet calculations included the Coulomb interactions at the nickel site parametrized as the direct Slater integrals F$^2_{dd}$ and F$^4_{dd}$ which were scaled to 80\% of the Hartree Fock values. The Coulomb interaction U$_{dd}$ of 5 eV was used according to photoemission experiments on similar materials \cite{Laan1986}. 
The charge-transfer energy $\Delta \approxeq$ 3.75 eV and the ligand-metal hybridization V$_{eg}$ and V$_{t2g}$ were optimized by comparison to known excitations from RIXS and optical data \cite{Kim2023}. We used the empirical relationV$_{eg} = \frac{3}{5} V_{t2g}$ and found that the energy level diagram reproduced the covalent behavior expected for the substituted NiPS$_3$ using the relations V$_{eg} = 0.184 \Delta + 1.301$ in accordance to previous work \cite{Occhialini2024}. The bare octahedral-symmetry Ni-3d crystal field splitting (`ionic') was fixed to Dq = 0.06 eV which leads to an effective crystal field splitting of 1 eV.
\label{CTM calculation}
\medskip
\\
\textbf{Supporting Information} \par

\section{Acknowledgements}
L.H acknowledges I-HUB, National Mission on Interdisciplinary CyberPhysical Systems (NM-ICPS) of the Department of Science and Technology, Government of India, for financial support. This work was supported by the Simons Foundation (Grant No. 1027114).  L.S and Y.L are very grateful to Jacques Peretti, Alistair Rowe and Claude Weisbuch for their support and contribution in the development of the experimental setup used for the optical measurements.

\bibliographystyle{apsrev4-2}

\bibliography{nips3.bib}

\newpage
\uppercase{Table of contents (TOC)}
\\

 \begin{center}
  \includegraphics[width=0.8\textwidth]{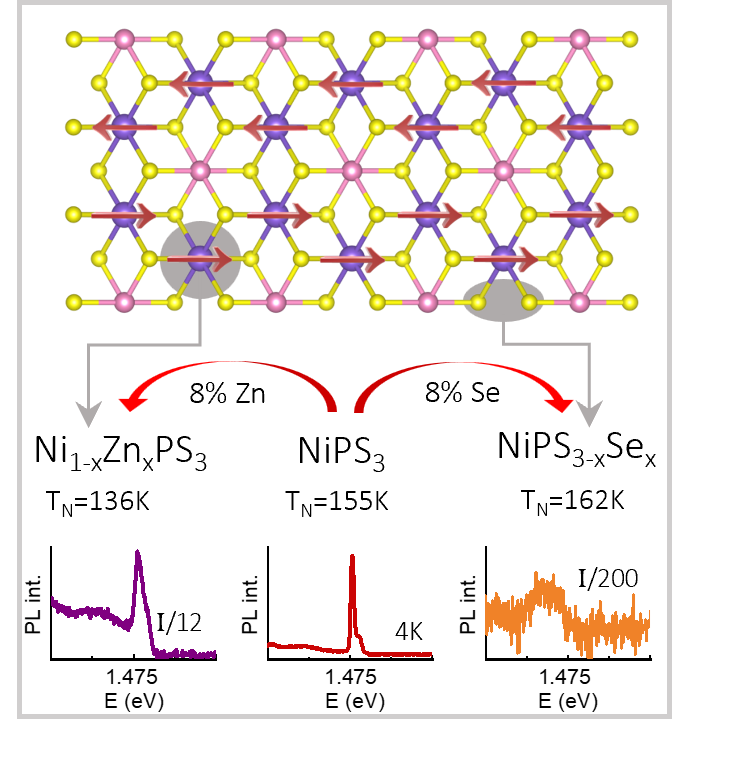}
   \end{center}
   Optical control and manipulation of magnetism is an issue of importance for future devices. Using metal and ligand substitution, this work connects sharp low-temperature photoluminescence in \nips to excited states of the Ni atom, to the flipping of its spin and to the tuning of the crystal field by its environment.
\end{document}